\documentclass{mem}
\usepackage{natbib}\usepackage{txfonts}\usepackage{balance}
\usepackage{graphicx}
\idline{??}{1}
\begin{document}
\def\teff{$T\rm_{eff }$}
\def\kms{$\mathrm {km s}^{-1}$}

\title{The Evolution of Bar Pattern Speed with Time and Bulge Prominence}


\author{Dimitri A. \,Gadotti}

\offprints{D. A. Gadotti}

\institute{Max Planck Institute for Astrophysics\\
Karl Schwarzschild Str. 1
D-85741 Garching bei Muenchen, Germany
}

\authorrunning{Gadotti}

\titlerunning{Evolution of Bar Pattern Speed}

\abstract{Results from the modelling of bars in nearly 300 galaxies are used to test predictions
from theoretical work on the evolution of bars. Correlations are found between bar ellipticity and boxiness,
between bar strength and normalised size, between the normalised sizes of bars and bulges, and between
normalised bar size and bulge-to-total ratio. Bars with different ellipticities follow
parallel lines in the latter two correlations. These correlations suggest that, formed with different
sizes and ellipticities, bars slow down and grow longer and stronger, in agreement with theoretical work.
As a consequence, bar pattern speeds should become lower with time, and towards galaxies with
more prominent bulges.
}
\maketitle{}

\section{Predictions from Theory}
\label{sec:intro}

A number of independent results from numerical simulations suggest that, in the case of galaxies with low
gas content, bars should slow down with time, as a result of angular momentum exchange from the inner
disc to the outer disc or halo. This slowdown allows bars to grow in size, capturing stars from the disc.
In fact, these simulations indicate that bars should also grow longer and stronger
\citep[see e.g.][and references therein]{AthMis02,DebMayCar06,MarShlHel06,BerShlJog06}. An analytical
treatment of these processes can be found in \citet{Ath03}. Observations suggest that the strong bar
in NGC 4608 has increased in mass by a factor of $\approx$1.7, through the capture of $\approx$13 per cent
of disc stars \citep{Gad08b}.

The purpose of this
work is to try to verify, using imaging data for a sample of nearly 300 barred galaxies, whether
these theoretical predictions are fulfilled. The effects of a high gas content and other difficulties
are briefly discussed in Sect. \ref{sec:dis}.

\section{Measuring the Properties of Bars from Observations}

\begin{figure*}
   \centering
   \includegraphics[keepaspectratio=true,width=4cm,clip=true]{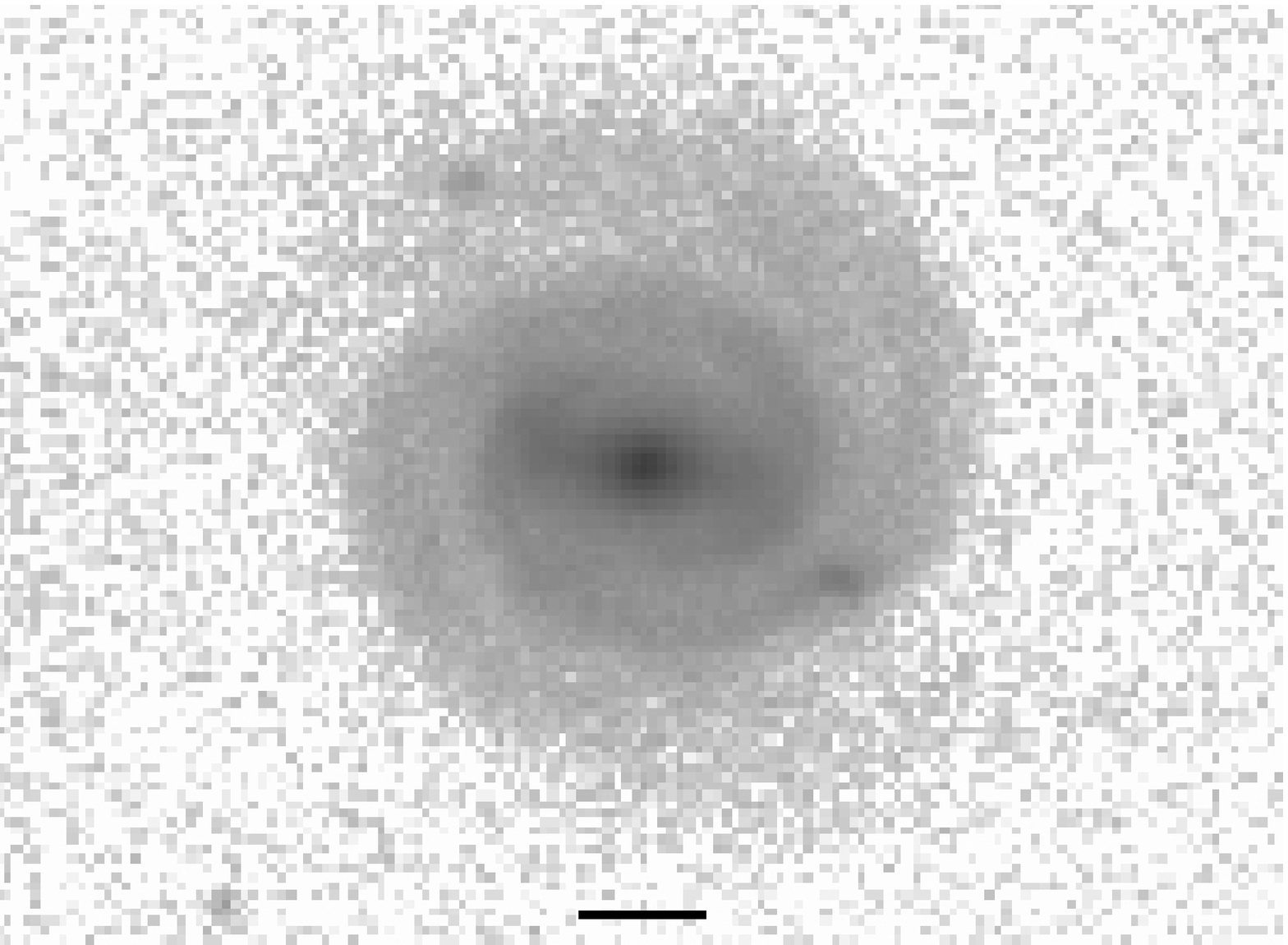}
   \includegraphics[keepaspectratio=true,width=4cm,clip=true]{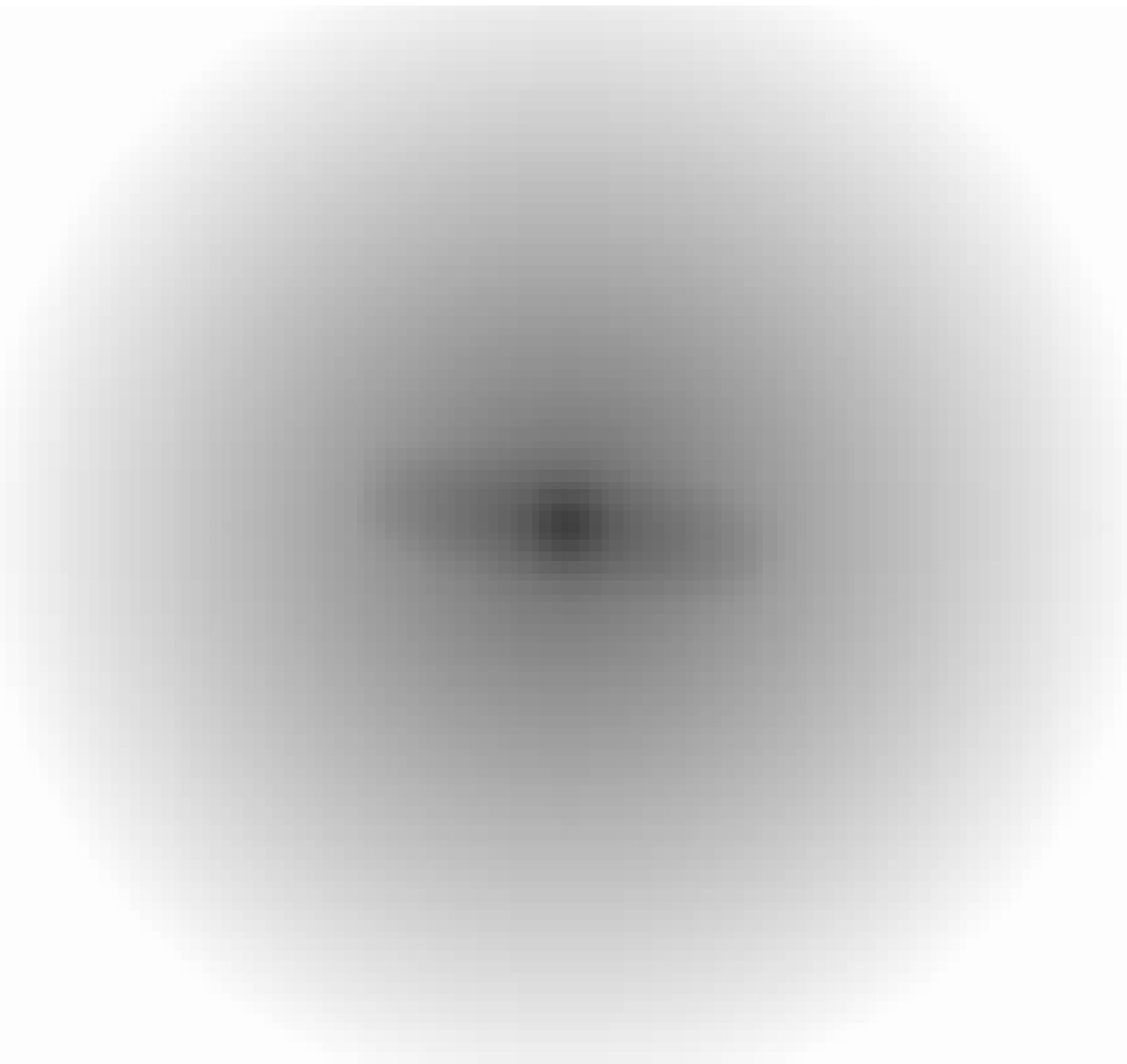}
   \includegraphics[keepaspectratio=true,width=4cm,clip=true]{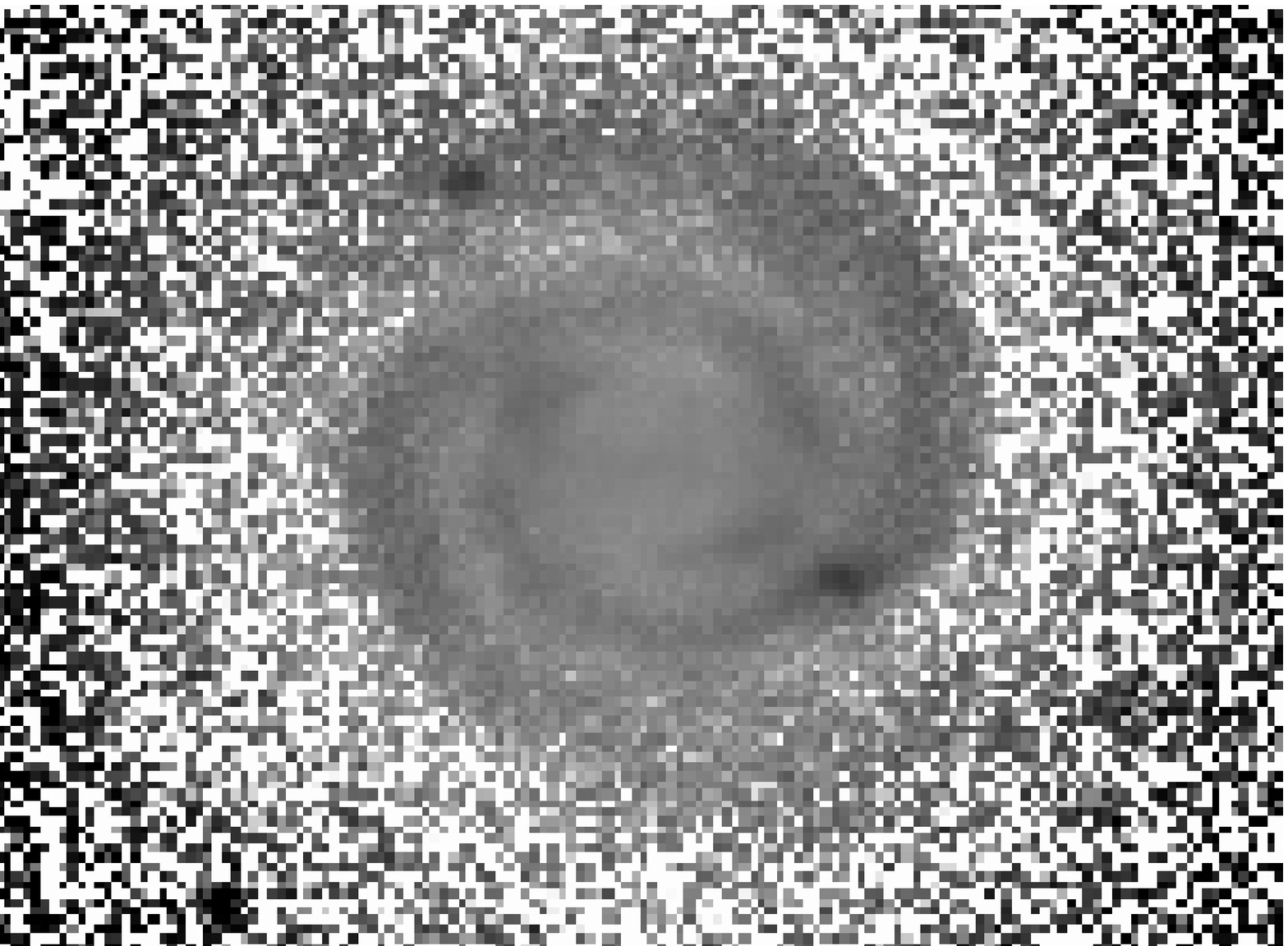}
   \includegraphics[keepaspectratio=true,width=6.5cm,clip=true]{gadotti1d.eps}
   \caption{Example of a bar/bulge/disc decomposition using {\sc budda} with one of the galaxies
in the sample. The top three panels show, from left to right, the original $i$-band galaxy image,
the model and residuals. The horizontal line in the galaxy image marks a length of 5 kpc.
The residual image is obtained subtracting the model
from the original image, and is displayed with a narrow intensity range,
in order to enhance residual sub-structures. In the residual image, darker pixels indicate where
the galaxy is brighter than the model, whereas whiter pixels indicate where the model is brighter than the galaxy.
Surface brightness profiles, obtained from cuts along the bar major axis, are shown in the bottom panels.
The dashed line corresponds to the original image, whereas the black solid line
corresponds to the total model. Red, blue and green lines refer,
respectively, to bulge, disc and bar. The dotted line in the lower panel shows the residuals (galaxy $-$ model).
{\it(Taken from \citealt{Gad08c}.)}}
   \label{fig:decomp}
\end{figure*}

In \citet{Gad08c}, the structural properties of nearly 1000 massive local galaxies in the Sloan Digital Sky Survey (SDSS)
were obtained through careful bar/bulge/disc decomposition, using the {\sc budda} code \citep{deSGaddos04,Gad08b}.
In this sample, there are 291 barred galaxies, and the code is able to estimate a number of structural parameters,
such as bulge effective radius $r_e$, bulge-to-total ratio $B/T$, bar effective radius $r_{e,{\rm bar}}$ and length
(semi-major axis) $L_{{\rm Bar}}$, bar ellipticity $\epsilon$, and boxiness $c$. An example of such decompositions
is shown in Fig. \ref{fig:decomp}.

The surface brightness profiles of bars are modelled with S\'ersic functions, as are those of bulges. However, while bulges
have S\'ersic indices $n$ typically in the range $1\lesssim n\lesssim6$, bars have S\'ersic indices $n_{{\rm bar}}$
typically in the range $0.5\lesssim n_{{\rm bar}}\lesssim1$.
In addition, bars are fitted as a set of concentric, generalised ellipses \citep[see][]{AthMorWoz90}:

\begin{equation}
\left(\frac{|x|}{a}\right)^c+\left(\frac{|y|}{b}\right)^c=1,
\end{equation}

\noindent where $a$ and $b$ are the bar semi-major and semi-minor axes, respectively, and $c$ is the bar
boxiness. If $c=2$, then the bar is a perfect ellipse, whereas if $c>2$ the bar has a more rectangular shape.
Since bars generally do have rectangular shapes, it is important to fit them with $c>2$. The results
presented here were obtained with $c$ as a free parameter.
The bar ellipticity is thus $1-b/a$.
As noted in \citet{Gad08b}, bar ellipticities obtained through
detailed image decomposition are more reliable than those obtained via ellipse fits to isophotes.
In the latter, the superposition of light from bulge and disc makes the isophotes in the
bar region rounder, leading to a systematic underestimation of the bar ellipticity.

It should also be noted that, due to the relatively poor spatial resolution of SDSS images, bars with
$L_{{\rm Bar}}\lesssim2-3$ kpc, typically seen in very late-type spirals
(later than Sc -- \citealt{ElmElm85}), are frequently missed.
Thus, the results presented here concern bonafide, large bars, as those typically seen in early-type
disc galaxies.

\section{Bar Strength and Growth}

\begin{figure}
   \centering
   \includegraphics[keepaspectratio=true,width=5.5cm,clip=true]{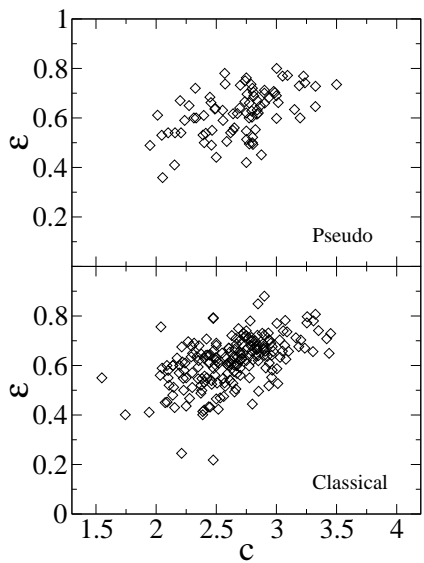}
   \caption{Bar ellipticity plotted against boxiness for galaxies with
classical and pseudo-bulges.}
   \label{fig:ec}
\end{figure}

Figure \ref{fig:ec} shows that bar ellipticity and boxiness are correlated, and this does not
depend on whether the galaxy hosts a classical bulge or a pseudo-bulge (see \citealt{Gad08c} for how
pseudo-bulges are identified). Furthermore, both parameters are related
to the strength of bar. Keeping everything else the same, a more eccentric (or more rectangular) bar
introduces a more substantial non-axisymmetric perturbation in the galaxy potential. Thus, one can use
the product $\epsilon\times c$ as a measure of bar strength.

\begin{figure}
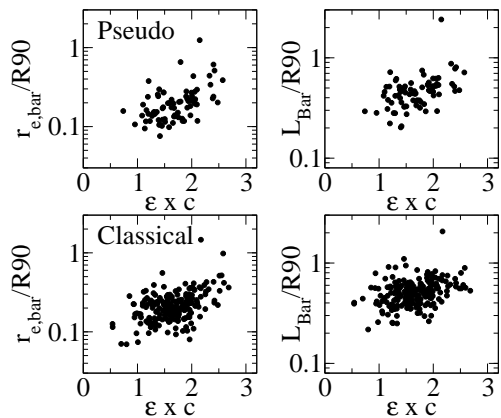

   \centering
   \includegraphics[keepaspectratio=true,width=6.5cm,clip=true]{gadotti3a.eps}\\
   \includegraphics[keepaspectratio=true,width=6.5cm,clip=true]{gadotti3b.eps}
   \caption{Normalised measures of bar size plotted against bar strength for galaxies
with classical and pseudo-bulges. Longer bars tend to be stronger.}
   \label{fig:barsl}
\end{figure}

Now, one can see if bar size correlates with bar strength (parameterised as $\epsilon\times c$),
as in the theoretical studies mentioned in Sect. \ref{sec:intro}. One can use either $r_{e,{\rm bar}}$ or
$L_{{\rm Bar}}$ as a measure of bar size. However, larger discs will form larger bars, and, in fact, bar size is
correlated with disc size, at least in early-type disc galaxies \citep[see e.g.][]{Erw05}. Therefore, bar size
must be normalised, and, with this aim, one can divide $r_{e,{\rm bar}}$ or $L_{{\rm Bar}}$ by the disc
scalelength $h$, or by the radius that contains 90 per cent of the galaxy light $R90$, or by the semi-major axis of the
24 $r$-band mag arcsec$^{-2}$ isophote $r_{24}$. One can obtain $h$ and $r_{24}$ from the {\sc budda} models,
while $R90$ is available in the SDSS database. The results presented here are essentially similar for all
such possible parameterisations of normalised bar size.
Figure \ref{fig:barsl} shows that longer bars tend to be stronger in galaxies with classical and pseudo-bulges,
in agreement with the theoretical work.

\begin{figure}
   \centering
   \includegraphics[keepaspectratio=true,width=6.5cm,clip=true]{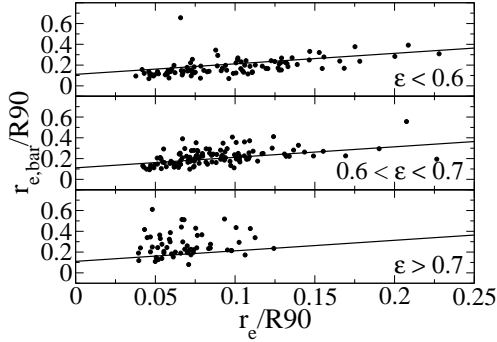}
   \caption{Correlation between the normalised sizes of bars and bulges,
for bars in three bins of ellipticity. The solid line shows a fit to all bars and is the same in
the three panels. Note that when $\epsilon<0.6$ most of the data points are below the line; when
$0.6<\epsilon<0.7$ the points follow the line more closely; and when $\epsilon>0.7$ most
of the points are above the line. Although there is no {\em clear} correlation for bars with
$\epsilon>0.7$ alone, bars with different ellipticities seem to describe parallel
lines in this correlation.}
   \label{fig:barbul}
\end{figure}

Figure \ref{fig:barbul} shows that the normalised sizes of bars and bulges are also correlated.
This figure confirms the correlation found by \citet{AthMar80} with a sample of 32 galaxies, and shows
that this correlation is present in a larger range of normalised bar and bulge sizes than previously found.
With the larger sample of the present study, a new aspect of the correlation is found, namely
that bars with different ellipticities seem to follow parallel tracks, although there is no
{\em clear} correlation for bars with $\epsilon>0.7$ alone.

\begin{figure}
   \centering
   \includegraphics[keepaspectratio=true,width=6.5cm,clip=true]{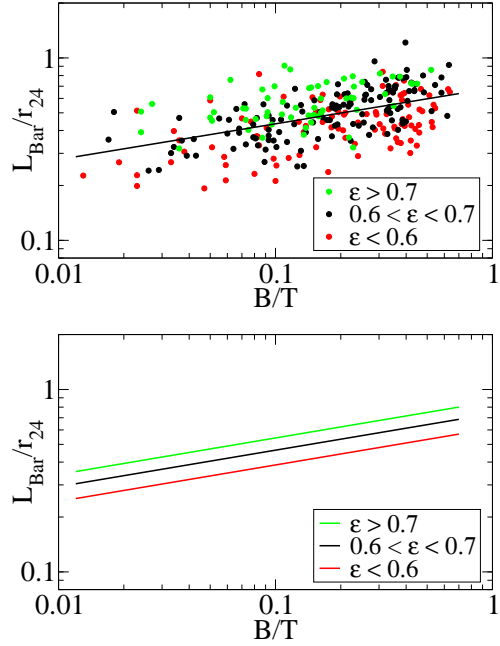}
   \caption{Normalised bar size plotted against bulge-to-total ratio, for bars in three bins of
ellipticity. The top panel shows the data points and the solid line is a fit to all points.
Longer bars tend to reside in galaxies with more conspicuous bulges. Furthermore,
bars with different ellipticities describe parallel lines in this relation. This is better seen in
the lower panel, which shows separate fits in the different ellipticity bins, with the slope fixed
at the value in the fit to all points.}
   \label{fig:barbtl}
\end{figure}

The correlation between the normalised sizes of bars and bulges suggests that the growth of
both components is somehow connected. Consistent with this idea, Fig. \ref{fig:barbtl} shows that
the normalised size of bars is also correlated with bulge-to-total ratio. Longer bars tend
to reside in galaxies with more conspicuous bulges. And, again, bars with different ellipticities
describe parallel lines in this relation. This is now more evident than in the relation
between the normalised sizes of bars and bulges (Fig. \ref{fig:barbul}). Furthermore, it is clear
that bars of all ellipticities, including those with $\epsilon>0.7$, follow a correlation
between their normalised sizes and $B/T$.

Galaxies with larger values of $B/T$ tend to be more massive \citep[see e.g.][]{Gad08c}.
In the current picture of galaxy formation, more massive galaxies are believed to form
earlier than less massive galaxies, as suggested by \citet{CowSonHu96}. In this case,
$B/T$ may serve as a proxy for time: galaxies with larger $B/T$ are older than those with
smaller $B/T$. If this is correct, then the correlation between normalised bar size and $B/T$
(Fig. \ref{fig:barbtl})
indicates that bars are longer in more evolved galaxies, and thus grow longer with time.
Combining this result with the correlation between normalised bar size and bar strength
(Fig. \ref{fig:barsl}), one concludes that bars grow longer and stronger with time, in
agreement with the theoretical predictions described in Sect. \ref{sec:intro}.

Using a different methodology, \citet{ElmElmKna07} also find that normalised bar size correlates
with bar strength and with the galaxy central density, and reach similar conclusions. \citet{SheElmElm08}
find that more massive galaxies have their bars in place at higher redshifts, whereas less massive
galaxies form bars at later times. They suggest that discs in more massive systems reach
a dynamical maturity earlier than those in less massive systems, and thus are able to form bars at earlier
times, in agreement with the results and interpretations given here.

The current understanding of the orbital structure in barred galaxies, and the observation that most bars end
near their corotation radius \citep[see e.g.][and references therein]{RauSalLau05},
tell us that bars generally cannot grow longer if they do
not slow down. Therefore, the results presented here also suggest that bars slow down with time, which is,
again, consistent with theory. Such a relation between the bar pattern speed $\Omega$ and $B/T$
is in the sense that bars rotate slower in galaxies with more prominent bulges, since these
galaxies have longer bars. This relation can also
be seen as a dependence of $\Omega$ on Hubble type, although there is some scatter in the relation between
$B/T$ and Hubble type \citep[e.g.][]{LauSalBut07,GraWor08}. Given that direct measurements of bar pattern
speed are difficult, particularly for late-type spirals, it is not surprising that no solid conclusions
can presently be drawn about a dependence of bar pattern speed on Hubble type, on direct observational
grounds \citep[see][]{GerKuiMer03,TreButSal07}.

An oustanding and unforeseen new result in Figs. \ref{fig:barbul} and \ref{fig:barbtl} is the existence
of parallel tracks in e.g. the correlation between bar normalised size and $B/T$, for bars with different
ellipticities. A straightforward way of interpreting the existence of these parallel tracks is to conceive
that bars should form with different normalised sizes and ellipticities, and then follow a somewhat parallel
growth. This is a new aspect that can be investigated with theoretical work. For instance, can
simulations form bars with various normalised sizes and ellipticities? Do simulations show that,
although formed with somewhat different properties, such bars follow a similar evolutionary path?

\section{Discussion}
\label{sec:dis}

The correlations presented above essentially corroborate the picture provided by theoretical
work on the formation and evolution of bars, at least when the galaxy gas content has little effect
on the bar. The role of gas is still a matter of debate \citep[see][]{BouComSem05,DebMayCar06,BerShlMar07}.
Nevertheless, most simulations show that the effect of gas is to {\em weaken} the bar with time.
Furthermore, in some gas-rich simulations, the bar strength oscillates substantially, alternating between
very weak and very strong in cycles. The evolution of the bar pattern speed seems not to be significantly
affected by the presence of gas. Bars still slow down in gas-rich simulations, which means they
can also grow longer.

Another issue that can complicate the interpretation of the observational results presented here
is related to the bar vertical buckling instability. Many simulations show that, at early stages, after
becoming very strong, bars grow vertically thick in their central parts, off the disc plane.
Due to this process, bars weaken dramatically, and then start growing stronger again. The strongest
buckling instability occurs soon after the formation of the bar, and ends when the bar is
$\approx2-3$ Gyr old.

It should also be mentioned that, depending on the initial conditions, the evolution of bar properties
in simulations can be very slow. All these issues might contribute to the scatter seen in the observed
correlations above. Nonetheless, since these correlations are based on the observation of typically
large bars, and since {\em short} bars are associated with gas-rich systems, it is likely that most of
the bars in the sample have not been significantly affected by the effects induced by gas.
Furthermore, since most long bars are not expected to have been recently formed
(this comes not only from simulations -- see \citealt{GaddeS06}), it is also likely
that most bars in the sample have already gone through the most significant vertical buckling.
If these suppositions are correct, then one indeed expects to see the correlations presented.
A similar work, with a sample of short bars, in gas-rich systems, is likely to shed light
on these issues.

Finally, it should be noted that the determination of the bar parameters involved in the results
above, namely bar length, ellipticity and boxiness, is difficult. The scatter observed might
at least partially result from measurement uncertainties (which are themselves also difficult to estimate).

\begin{acknowledgements}
I am grateful to the organisers for their efforts in making this a very fruitful meeting.
I thank Guinevere Kauffmann for her support throughout this work and useful discussions, and
Lia Athanassoula and Peter Erwin for important suggestions.
DAG is supported by the Deutsche Forschungsgemeinschaft and the Max Planck Society. Funding
for the SDSS project has been provided by the Alfred P. Sloan Foundation,
the SDSS member institutions, the National Aeronautics and Space
Administration, the National Science Foundation, the US Department
of Energy, the Japanese Monbukagakusho, and the Max Planck
Society.
\end{acknowledgements}

\bibliographystyle{aa}
\bibliography{/home/dimitri/papers/gadotti_refs}

\begin{thebibliography}{23}
\expandafter\ifx\csname natexlab\endcsname\relax\def\natexlab#1{#1}\fi

\bibitem[{{Athanassoula}(2003)}]{Ath03}
{Athanassoula}, E. 2003, \mnras, 341, 1179

\bibitem[{{Athanassoula} \& {Martinet}(1980)}]{AthMar80}
{Athanassoula}, E. \& {Martinet}, L. 1980, \aap, 87, L10

\bibitem[{{Athanassoula} \& {Misiriotis}(2002)}]{AthMis02}
{Athanassoula}, E. \& {Misiriotis}, A. 2002, \mnras, 330, 35

\bibitem[{{Athanassoula} {et~al.}(1990){Athanassoula}, {Morin}, {Wozniak},
  {Puy}, {Pierce}, {Lombard}, \& {Bosma}}]{AthMorWoz90}
{Athanassoula}, E., {Morin}, S., {Wozniak}, H., {et~al.} 1990, \mnras, 245, 130

\bibitem[{{Berentzen} {et~al.}(2006){Berentzen}, {Shlosman}, \&
  {Jogee}}]{BerShlJog06}
{Berentzen}, I., {Shlosman}, I., \& {Jogee}, S. 2006, \apj, 637, 582

\bibitem[{{Berentzen} {et~al.}(2007){Berentzen}, {Shlosman},
  {Martinez-Valpuesta}, \& {Heller}}]{BerShlMar07}
{Berentzen}, I., {Shlosman}, I., {Martinez-Valpuesta}, I., \& {Heller}, C.~H.
  2007, \apj, 666, 189

\bibitem[{{Bournaud} {et~al.}(2005){Bournaud}, {Combes}, \&
  {Semelin}}]{BouComSem05}
{Bournaud}, F., {Combes}, F., \& {Semelin}, B. 2005, \mnras, 364, L18

\bibitem[{{Cowie} {et~al.}(1996){Cowie}, {Songaila}, {Hu}, \&
  {Cohen}}]{CowSonHu96}
{Cowie}, L.~L., {Songaila}, A., {Hu}, E.~M., \& {Cohen}, J.~G. 1996, \aj, 112,
  839

\bibitem[{{de Souza} {et~al.}(2004){de Souza}, {Gadotti}, \& {dos
  Anjos}}]{deSGaddos04}
{de Souza}, R.~E., {Gadotti}, D.~A., \& {dos Anjos}, S. 2004, \apjs, 153, 411

\bibitem[{{Debattista} {et~al.}(2006){Debattista}, {Mayer}, {Carollo}, {Moore},
  {Wadsley}, \& {Quinn}}]{DebMayCar06}
{Debattista}, V.~P., {Mayer}, L., {Carollo}, C.~M., {et~al.} 2006, \apj, 645,
  209

\bibitem[{{Elmegreen} \& {Elmegreen}(1985)}]{ElmElm85}
{Elmegreen}, B.~G. \& {Elmegreen}, D.~M. 1985, \apj, 288, 438

\bibitem[{{Elmegreen} {et~al.}(2007){Elmegreen}, {Elmegreen}, {Knapen}, {Buta},
  {Block}, \& {Puerari}}]{ElmElmKna07}
{Elmegreen}, B.~G., {Elmegreen}, D.~M., {Knapen}, J.~H., {et~al.} 2007, \apjl,
  670, L97

\bibitem[{{Erwin}(2005)}]{Erw05}
{Erwin}, P. 2005, \mnras, 364, 283

\bibitem[{{Gadotti}(2008{\natexlab{a}})}]{Gad08b}
{Gadotti}, D.~A. 2008{\natexlab{a}}, \mnras, 384, 420

\bibitem[{{Gadotti}(2008{\natexlab{b}})}]{Gad08c}
{Gadotti}, D.~A. 2008{\natexlab{b}}, \mnras, in press, arXiv:0810.1953

\bibitem[{{Gadotti} \& {de Souza}(2006)}]{GaddeS06}
{Gadotti}, D.~A. \& {de Souza}, R.~E. 2006, \apjs, 163, 270

\bibitem[{{Gerssen} {et~al.}(2003){Gerssen}, {Kuijken}, \&
  {Merrifield}}]{GerKuiMer03}
{Gerssen}, J., {Kuijken}, K., \& {Merrifield}, M.~R. 2003, \mnras, 345, 261

\bibitem[{{Graham} \& {Worley}(2008)}]{GraWor08}
{Graham}, A.~W. \& {Worley}, C.~C. 2008, \mnras, 388, 1708

\bibitem[{{Laurikainen} {et~al.}(2007){Laurikainen}, {Salo}, {Buta}, \&
  {Knapen}}]{LauSalBut07}
{Laurikainen}, E., {Salo}, H., {Buta}, R., \& {Knapen}, J.~H. 2007, \mnras,
  381, 401

\bibitem[{{Martinez-Valpuesta} {et~al.}(2006){Martinez-Valpuesta}, {Shlosman},
  \& {Heller}}]{MarShlHel06}
{Martinez-Valpuesta}, I., {Shlosman}, I., \& {Heller}, C. 2006, \apj, 637, 214

\bibitem[{{Rautiainen} {et~al.}(2005){Rautiainen}, {Salo}, \&
  {Laurikainen}}]{RauSalLau05}
{Rautiainen}, P., {Salo}, H., \& {Laurikainen}, E. 2005, \apjl, 631, L129

\bibitem[{{Sheth} {et~al.}(2008){Sheth}, {Elmegreen}, {Elmegreen}, {Capak},
  {Abraham}, {Athanassoula}, {Ellis}, {Mobasher}, {Salvato}, {Schinnerer},
  {Scoville}, {Spalsbury}, {Strubbe}, {Carollo}, {Rich}, \&
  {West}}]{SheElmElm08}
{Sheth}, K., {Elmegreen}, D.~M., {Elmegreen}, B.~G., {et~al.} 2008, \apj, 675,
  1141

\bibitem[{{Treuthardt} {et~al.}(2007){Treuthardt}, {Buta}, {Salo}, \&
  {Laurikainen}}]{TreButSal07}
{Treuthardt}, P., {Buta}, R., {Salo}, H., \& {Laurikainen}, E. 2007, \aj, 134,
  1195

\end{thebibliography}

\end{document}